# Investigation into the nature behind the interesting half levitation behavior of claimed superconductor LK-99

Lingyi Liao[1], Zihao Chen[1], Yuanyuan Tan[1], Qingsong Mei[1, *]

[1] School of Power and Mechanical Engineering, Wuhan University, Wuhan, 430072, China

*Corresponding author. School of Power and Mechanical Engineering, Wuhan University, Wuhan, 430072, China. E-mail address: qsmei@whu.edu.cn (Q.S. Mei).

## Abstract

A recent article published by Lee et.al. claimed to have successfully achieved superconductivity at room temperature (RT) has become a topical issue. Besides the research paper, Lee and his team provided a demonstration video of LK-99 half levitating (HL) on a magnet. Such interesting HL appearance has drawn tremendous sensation both in academia and the network. However, the true identity of LK-99 still remains unclear, i.e., whether the HL behavior can necessarily indicate the diamagnetism behavior of the sample. Here, we fabricated our own LK-99 samples following the procedures reported by Lee et al. We found quite a few sample pieces showing the typical HL that is similar to those reported. Meanwhile, oxidation during the sample preparation was found to deleterious to acquiring HL in the sample, while furnace cooling or water quenching in the last step revealed little effect. However, our careful observations indicated that those HL pieces are more likely simple ferromagnetic. Then we conducted a comprehensive study on the behavior patterns of typical diamagnetism and ferromagnetic substances interacting with a $Nd_2Fe_{14}B$ magnet, and provided instructions to distinguish the characteristics between ferromagnetic and diamagnetic to prevent misunderstanding of LK-99 like levitation behavior.

Key words: LK-99, ferromagnetic, diamagnetic, half levitation

# 1. Introduction

Superconductor is one of the most famous materials since it has been discovered. Superconductivity represents a material reveals none resistance and full diamagnetism, once activated, a superconductor could realize zero electric resistance and the heat loss of current flow would be completely avoid, thus creating a powerful magnetic field. Unfortunately, superconductivity could only be achieved under extreme conditions, which demands low temperature and high pressure. Such extreme condition heavily restricts the application of superconductors, and all researchers are keen to discover high temperature superconductors with a critical temperature above 0°C (273.15 K). Currently, Bi based, Y based superconductors are most practical and promising candidates, with a critical temperature above 92 K and 110 K [1], and apparently it is an arduous journey and requires enormous effort to achieve high critical temperature, even possibly, at RT [2].

Remarkably, Lee et.al.[3, 4] announced to have successfully synthesized RT superconductor at ambient pressure with a modified lead-apatite (LK-99) structure. In the published paper, they claimed to have observed zero resistivity and Meissner effect, most importantly, they provided a demonstration video presenting a LK-99 sample HL on a permanent magnet. Among these many evidences in the effort to prove the validity of superconductivity of LK-99, the demonstration of Meissner effect through putting the sample on a permanent magnet is most concise and explicit, likewise, the provided video has received the most intense discussion. The interesting HL behavior of LK-99 seems to be quite persuasive evidence demonstrating diamagnetism, considering ferromagnetic substance would get attracted to the magnet in normal cognition.

Such findings have instantly caused worldwide attention, and immediate reproduction of LK-99 has been attempted by several different institutions, following the synthesis instructions provided in by Lee et.al [5-11]. However, the replication of LK-99 turned out to be quite tricky, although a few theoretical analysis reported that the strategy of LK-99 is promising [5, 6, 9] and abnormal resistance temperature behavior has been observed [11], the positive results seemed to be confined within theoretical analysis. Meissner effect was not observed in any replicated samples, and the LK-99-like sample fabricated by Guo et.al [8] and Kumar et.al [12] were found out as ferromagnetism. Despite several replicated samples revealed similar HL behavior to the demonstration video provided by Lee and his team (Figure 1) [8, 13], the following test to provide convincing evidence for the superconductivity of LK-99 fall short. The editors from Nature briefly summarized the replication attempts and sentenced that LK-99 is not a superconductor [14, 15].

Therefore, we synthesized our own LK-99 sample following the same sintering procedure, and surprisingly, we managed to acquire several pieces of sample that revealed the similar HL behavior in the demonstration video, and the X-ray diffraction (XRD) analysis of the fabricated sample revealed similar pattern compared to the results of Lee et.al [3]. However, as we continue to investigate the magnetism of our LK-99 sample, it appeared to be ferromagnetic instead of diamagnetic, as it could be directly attracted when the magnet gradually approximated the sample from above. Consequently, such interesting HL behavior does not necessarily equal to diamagnetism, and the demonstration video does not provide sufficient scientific evidence for the diamagnetism of LK-99. To fully understand the levitation behavior and distinguish the magnetic substance between ferromagnetism and diamagnetism, we further conduct a comprehensive study on the behavior pattern of magnetic substance interacting with a permanent magnet. This paper investigated the behavior of both ferromagnetic and diamagnetic substance under magnetic field, we yield conclusion that simple levitation behavior does not support diamagnetism nor ferromagnetism of sample, and provide instructions to distinguish the characteristic of a magnetic substance between ferromagnetic and diamagnetic to prevent misunderstanding of LK-99 like HL behavior in the future.

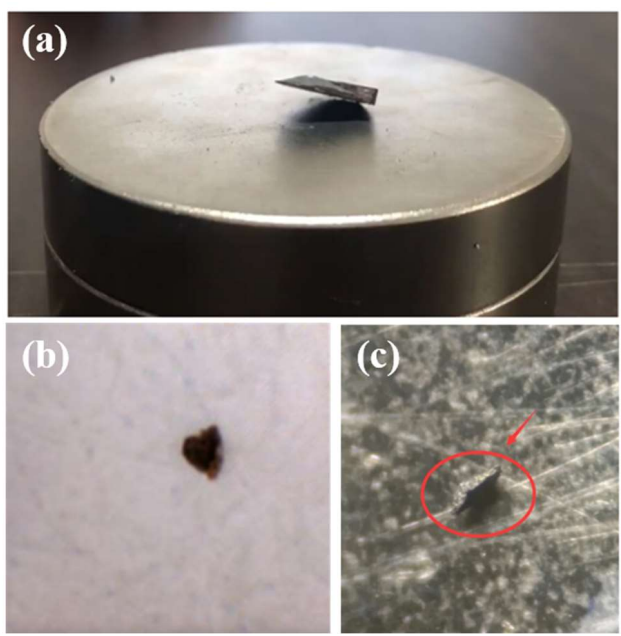

Figure 1. Optical images of HL behavior of synthesized samples by different institutions, (a) LK-99

by Lee et.al [3], (b) small piece fabricated by Wu et.al [13], (c) small piece fabricated by Guo et.al [8].

## 2. Experimental

**Sample preparation**

LK-99 ( $Pb_{10-x}Cu_x(PO_4)_6O$ (0.9 < $x$ < 1.1) ) were prepared with solid-state sintering method reported by Lee et.al [3]. The raw materials used in the process were PbO, $PbSO_4$, Cu and P (purchased from Shanghai Aladdin Biochemical Technology Co., Ltd). In the third step of LK-99 preparation, we set up two control tests to investigate the effects of water quenching and oxidation, detailed conditions are provided in Table 1. As a result, no significant difference in appearance was discovered between water quenching and furnace cooling. The unoxidized sample present dark-gray appearance, while the oxidized sample is yellow-gray. Phase identification of fabricated samples and intermedia products have been investigated by XRD (SmartLab SE) with Cu Kα radiation under 40 kV and 50 mA, with scanning range (2θ) from 10° to 90°, the step size of 0.01° and the scan speed of 15°/min.

Table 1. Different terms of conditions in the third step of processing process

| 1 | 2 | 3 | 4 |
| --- | --- | --- | --- |
| Water quenching | Water quenching | Furnace cooling | Furnace cooling |
| Sealed | Unsealed | Sealed | Unsealed |

In order to verify whether the sample has the half-levitation reported by Lee et.al, we put the sample on a piece of paper above the magnet, and let the $Nd_2Fe_{14}B$ magnet gradually approximate the sample then move away. Due to the large samples remained stationary, they were grounded to make small pieces in a crucible. Then we found a few small pieces which could achieve HL on the magnet in dark-grey samples. We picked out a relatively large piece from the samples for further investigations.

## 3. Results

## 3.1 XRD analysis

Figure 2 (a-c) shows the XRD patterns of fabricated LK-99 sample and intermedia products. The results of two intermedia products are in good correspondence with the target product, and the position of the peaks is almost identical compared with those reported by Lee et.al [3]. The appearance of the intermedia products and fabricated LK-99 sample were presented in Figure 2(d-g), the oxidized sample presented yellow-grey and the LK-99 sample without oxidation were dark-grey which corresponded to the LK-99 sample reported by Lee et.al. Through careful investigation, magnetic pieces were found only in dark-grey samples which were sealed to prevent oxidization. However, the oxidized yellow-grey sample failed to present any magnetism, and it is reasonable to assume that oxidation is deleterious to the production process.

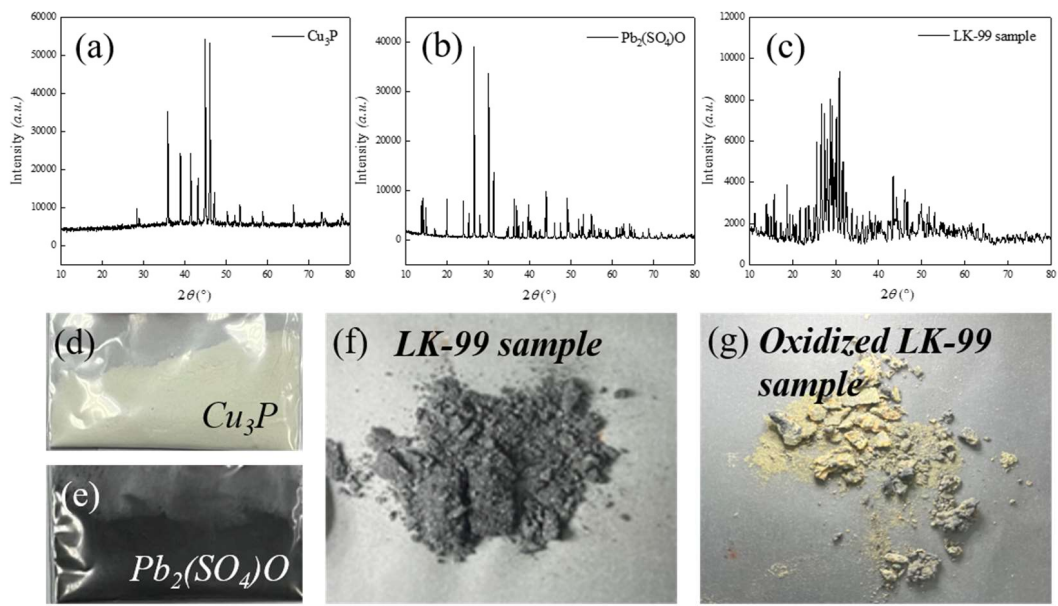

Figure 2. XRD patterns and optical images of fabricated LK-99 samples and intermedia products, (a) $Cu_3P$, (b) $Pb_2(SO_4)O$, (c) $Pb_{10-x}Cu_x(PO_4)_6O$ (0.9 < $x$ < 1.1), (d) $Pb_2(SO_4)O$, (e) $Cu_3P$, (f) LK-99 sample, (g) oxidized LK-99 sample.

## 3.2 Magnetic analysis

To fully investigate the behavior patterns of ferromagnetic and diamagnetic materials interacting with permanent magnet, we choose different samples to provide comprehensive demonstration (Figure 3), including the fabricated sample, iron piece (ferromagnetic), graphite

(diamagnetic) and pyrolytic graphite (diamagnetic). We simulated the interaction behavior demonstrated in other researches, through putting the magnetic sample on a piece of paper above the magnet, and let the magnet gradually approximate the sample then move away. The different magnetic samples were named from S1 to S4 to provide a more concise description as demonstrated in Table 2.

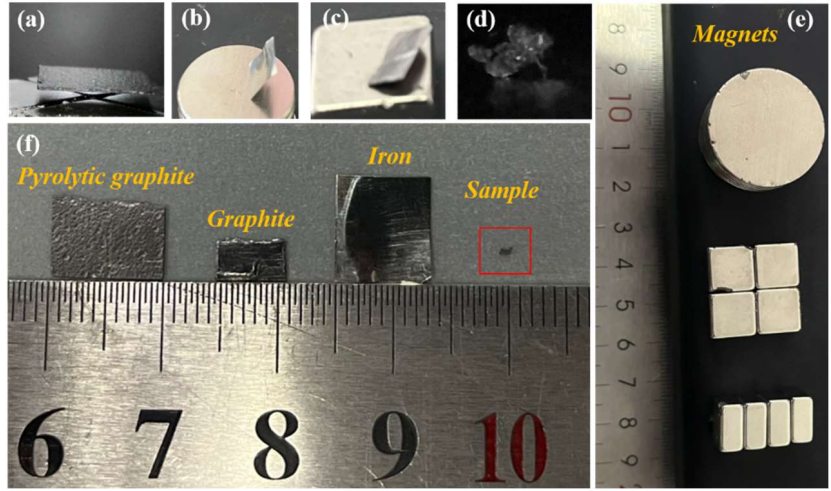

Figure 3. Optical images of samples and permanent magnets used in this study, (a-d) levitation of different samples on permanent magnets, (e) permanent magnets, (f) magnetic samples.

Table 2. Designation of different magnetic samples.

| S1 | S2 | S3 | S4 |
|---|---|---|---|
| LK-99 sample | Iron | Graphite | Pyrolytic graphite |
| To be determined | Ferromagnetic | Diamagnetic | Diamagnetic |

Figure 4 presents the HL behavior of S1 to S3, and surprisingly, it seems that ferromagnetic materials (S2) and diamagnetic materials (S3) revealed the same behavior pattern. As the magnet move beneath the sample from left to right, the far-end of sample always got lifted and the angle $\theta$ gradually increased. When the magnet is right beneath the sample, $\theta$ is near 90° and the sample shown vertical standing on the magnet. When the magnet continued to move away, $\theta$ continued to increase and the sample got flipped over from right to left, which is opposite to the motion direction of the magnet. Moreover, the fabricated LK-99 sample revealed same behavior patterns thus making the true identity of S1 more mysterious. Till this point, all samples presented HL on the magnet and pointed outwards the magnet alongside the direction of magnetic flux, similar to the commonly known phenomena that a magnet needle always points to north in a compass. More importantly, the angle $\theta$ appeared to be highly corresponded to the location of the sample on the magnet, as the sample presented same HL behavior pattern as the magnet approach and move away.

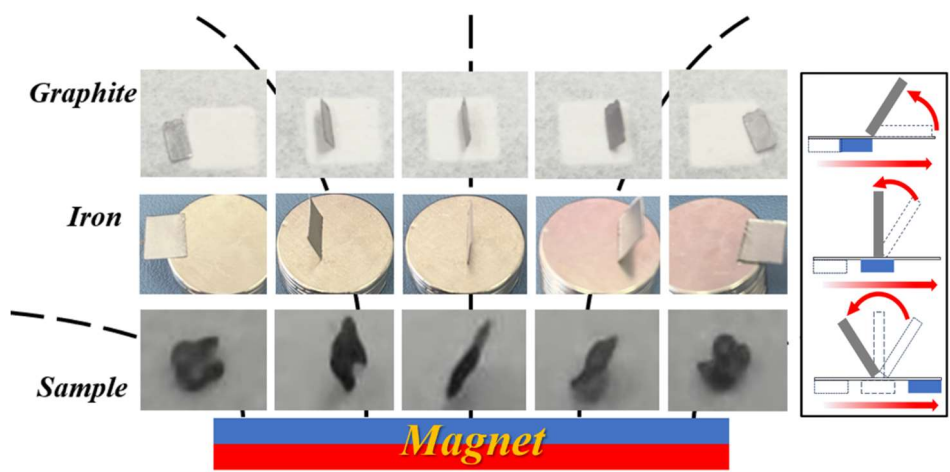

Figure 4. Behavior pattern of S1 to S3 on a permanent magnet.

To disclose the real magnetism of S1, we flipped the magnet over, in which case the ferromagnetic substance would still get attracted and diamagnetic substance would definitely get repelled to fall. Our results are shown in Figure 5 (a-c), like iron pieces, S1 could stand upside down on the bottom side of the magnet against gravity (G). It turned out that S1 is ferromagnetic and further evidence is provided as S1 could be attracted by the magnet when approaching from above. All these typical ferromagnetic phenomena confirmed that S1 is not diamagnetic. In addition, diamagnetic substance possesses its unique characteristics, such as full levitation on proper magnet fields. As demonstrated in Figure 5 (d-e), both S3 and S4 could achieve full levitation above four powerful permanent magnets aligned in square. The levitation height of S4 was higher than S3 due to the diamagnetism of pyrolytic graphite was stronger than graphite, and received larger repelling force in the magnet field.

Consequently, S1 has been identified as ferromagnetic, and more importantly, the HL behavior have been observed both in ferromagnetic and diamagnetic substances, which were failed to be noticed in previous study. Under these circumstances, the demonstration video provided by Lee et.al [3] seemed less persuasive and the true nature behind this interesting HL behavior still remained unclear. Therefore, we further conduct force analysis to investigate the mechanism of levitation behavior of magnetic substance.

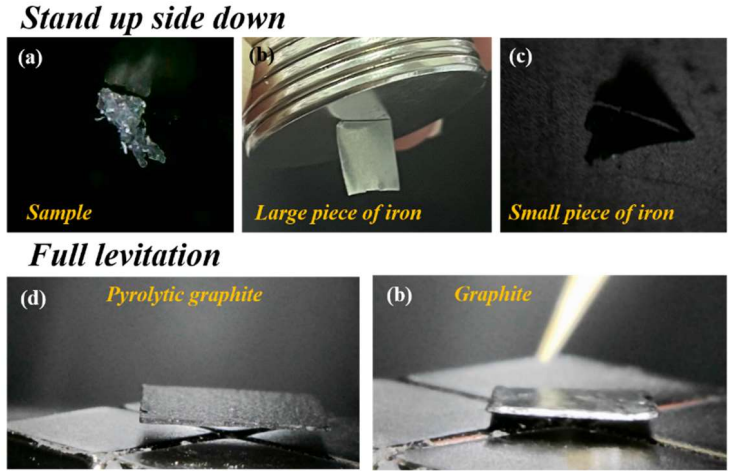

Figure 5. Different characteristic between ferromagnetic and diamagnetic substances. (a-c) levitation behavior of S1 and S2 when the magnet is flipped over, (d-e) full levitation of S3 and S4 above proper magnet fields.

## 4. Discussion

It is a common fact that any magnetic substance would first get magnetized and subjected to extra magnetic force once put into a magnetic field. Under ideal condition that no extra force is applied, the magnetic substance would rotate under the magnet torque and reach a balance position alongside the direction of magnetic force or magnetic flux. The magnetic force ($F_{magnet}$) is determined by the magnetic moment of the substance (m) and the magnetic field intensity (B), and $F_{magnet}$ is described as:

$$F_{magnet} = \nabla (m \times B) \quad (1)$$

where $\nabla$ represents the intensity of m and B in different vectors, and for a magnetic field of permanent magnet, B is heterogenous.

Therefore, when a homogenic magnetic substance considered as regular shape is put into the magnetic field of a permanent magnet, the force subjected to the substance differs from one end to another. Here, we consider the magnetic samples are all homogenic and the magnet field generated by the permanent magnet is gradient distributed. The force analysis of ferromagnetic and diamagnetic samples is shown in Figure 6. For both ferromagnetic and diamagnetic sample, once force could be balanced, the HL status could be achieved. Then we conduct brief force analysis, following the principle that the force in horizontal direction and vertical direction should balance, and the force torque of magnet force and G should balance as well.

Horizontal direction:

$$\sum F_x = F_a cos\theta + F_r cos\theta + F_f = 0 \quad (2)$$

Vertical direction:

$$\sum F_y = F_a sin\theta + F_r sin\theta + G + F_s = 0 \quad (3)$$

Torque:

$$\sum \tau = \tau(F_{magnet}) + \tau(G) = 0 \quad (4)$$

$$\sum \tau(F_{magnet}) = \tau(F_a) + \tau(F_r) = F_a sin\theta L_1 + F_r sin\theta L_2 \quad (5)$$

where $F_a$ stands for the attraction force, $F_r$ represents the repelling force, $F_f$ is the friction force, $F_s$ is the supporting force, $L_1$ and $L_2$ refers to the arm of force of $F_a$ and $F_r$, respectively. Considering the gradient distribution of magnetic field and the shape of sample, we provide following assumptions:

For ferromagnetic substance, $L_1 < L_2$ and $F_a > F_r$;

For diamagnetic substance, $L_1 > L_2$ and $F_a < F_r$.

The HL behavior already suggest that equation (2) and (3) could be satisfied, therefore once equation (4) is equaled, the conditions of balancing at HL could be achieved. As a matter of fact, the direction of $\tau(G)$ always points to earth and accordingly, the direction of $\tau(F_{magnet})$ must against $\tau(G)$ to maintain balance. Let's assume $\tau(G) < 0$, then $\tau(F_{magnet})$ should always be greater than 0. For both ferromagnetic and diamagnetic substances, the direction of $F_a$ and $F_r$ are opposite, and $sin\theta$ could be taken out as a common factor, thus leaving the variable between the arm of force, $L_1$ and $L_2$. Regard to our previous terms for ferromagnetism and diamagnetism, it seems that the balance is always achievable under proper conditions between the value of G and magnetic force. Additionally, when the HL sample was applied with an extra force, the balance would not be easily broke, suggesting that the existence of strong magnetic torque continuing to resist outer disturbance to maintain the initial balanced state, further demonstration for this resistivity is provided in Supplementary video 1-3. This could be explained by the lowest energy principle that once a balance status is achieved, the system prefers to maintain this balance state and it requires extra energy to break to existing balance and obtain a new balance condition. In this case, the magnetic torque played the role to keep balance and cause the HL behavior.

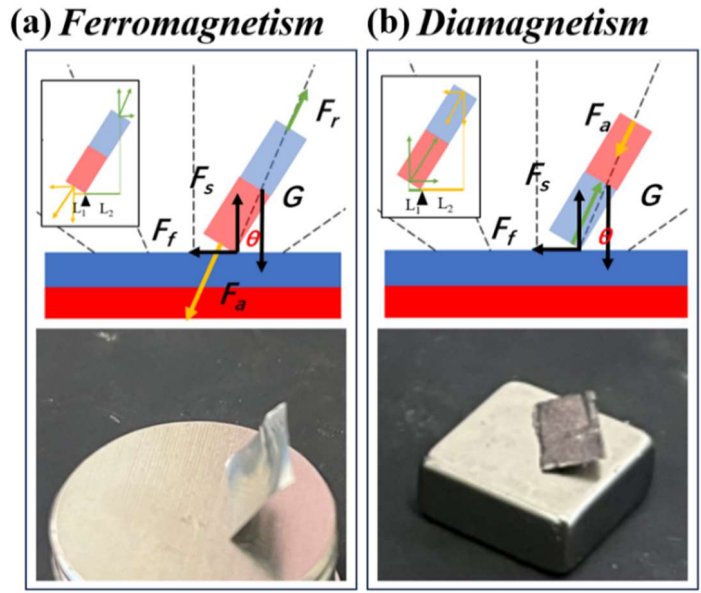

Figure 6. The force analysis of HL for S2 and S3.

In our perspective, the HL behavior is achievable in both ferromagnetic and diamagnetic substances. It is a matter of balancing between G and magnetic force. To further prove the validity of our analysis, we first conducted the flipped experiment again with S2 and observed the small change of angle $\theta$, as shown in Figure 7. Apparently, when the permanent magnet was flipped over, S2 could stand upside down on the bottom of the magnet, while the angle $\theta$ increased slightly from 68° to 75°. Such findings suggested that when the vector of $\tau(G)$ was reversed, the magnetic torque had to provide extra torque to balance $\tau(G)$, causing variation of $\theta$. Similar to previous discussion upon disturbing the sample with extra force, this upside down standing also represented resistivity to extra weight which is previously balanced by supporting force. Meanwhile, the angle $\theta$ seems to be a less important factor in verification of ferromagnetism and diamagnetism, and contradicted to our previous results presented in Figure 4, all S1 to S3 revealed the same behavior pattern on the permanent magnet. According to our analysis, provided suitable G, the HL could be achieved at different $\theta$ at same location on the magnet. As shown in Figure 8 (a-b), we changed the weight and center of G of S2 and S3 with a foam tape to alter the balance between $\tau(G)$ and $\tau(F_{magnet})$. And we successfully realized opposite HL direction (pointing inwards) at same location on the magnet. To be noted, the foam tape we used did not affect the contact between sample and the magnet, as demonstrate in the insert schematic picture of Figure 8, all foam tape was not adhered to the lower part of the sample. Apart from varying G, we adjust $\tau(F_{magnet})$ by using different magnetic fields composed of magnets (Figure 8 (c-d)), as demonstrated HL pointing inwards and outwards at same location on the magnet could be observed in S3 (Figure 3).

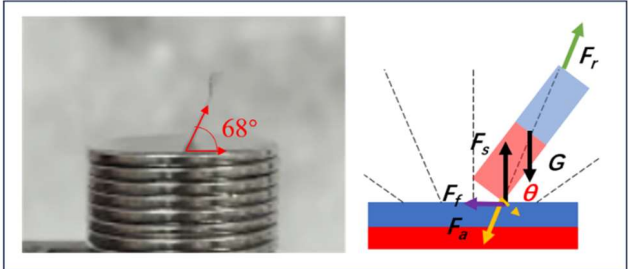

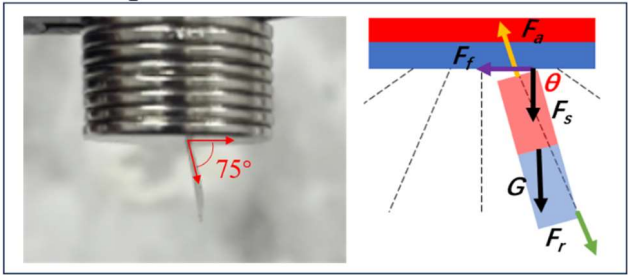

Figure 7. Variation of angle $\theta$ of ferromagnetic sample when the magnet is flipped over.

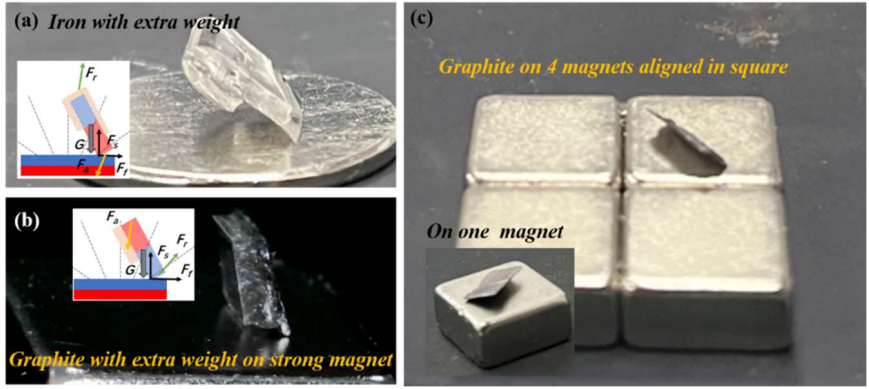

Figure 8. Demonstration for the feasibility of different angle when magnetic sample present HL on the magnet.

## 5. Conclusion

We have synthesized our own LK-99 samples following the provided procedure, and acquired ferromagnetic samples which also presented the interesting HL behavior. We thoroughly investigated the behavior of magnetic substance interacting with a permanent magnet and came to a conclusion that both ferromagnetic and diamagnetic substances could achieve HL on a permanent magnet. The effects of oxidation and cooling methods during the sample preparation were also investigated. The main conclusions could be drawn as follows:

(1) LK-99 samples have been fabricated and a few small pieces were found to show the HL behavior. However, our careful observations indicated that those pieces are more likely simple ferromagnetic. The oxidation during the sample preparation was found to have significant effect on achieving the ferromagnetic pieces. Water quenching on the last step of sample preparation was found to yield similar effect with furnace cooling.

(2) More importantly, we showed that both ferromagnetic or diamagnetic materials can show the interesting HL behavior, and the levitation behavior can be further changed by the inhomogeneity of the substances in the sample.

## Supplementary Materials

S1: https://www.capcut.cn/share/7275968102041425204?t=1

S2: https://www.capcut.cn/share/7275964428154197288?t=1

S3: https://www.capcut.cn/share/7275964428154197288?t=1

## Reference

[1] L. Liu, P. Yang, Z. Huang, Research progress in high - temperature superconducting, wire （tape）materials, Cryogenics and Superconductivity, 34 (2006) 48-51,67.
[2] W.E. Pickett, Colloquium: Room temperature superconductivity: The roles of theory and materials design, Reviews of Modern Physics, 95 (2023).
[3] S. Lee, J.-H. Kim, Y.-W. Kwon, The First Room-Temperature Ambient-Pressure Superconductor, Arxiv, (2023).
[4] S. Lee, J. Kim, S. Im, S. An, Y.-W. Kwon, K.H. Auh, Consideration for the development of room-temperature ambient-pressure superconductor (LK-99) (vol 33, pg 61, 2023), Journal of the Korean Crystal Growth and Crystal Technology, 33 (2023) 124-124.
[5] P. Abramian, A. Kuzanyan, V. Nikoghosyan, S. Teknowijoyo, A. Gulian, Some remarks on possible superconductivity of composition $Pb_9CuP_6O_{25}$, Arxiv, (2023).

[6] G. Baskaran, Broad Band Mott Localization is all you need for Hot Superconductivity: Atom Mott Insulator Theory for Cu-Pb Apatite, Arxiv, (2023).

[7] W. Chen, Berry curvature and quantum metric in copper-substituted lead phosphate apatite, Arxiv, (2023).

[8] K. Guo, Y. Li, S. Jia, Ferromagnetic half levitation of LK-99-like synthetic samples, Science China-Physics Mechanics & Astronomy, 66 (2023).

[9] L.Y. Hao, E.G. Fu, First-principles calculation on the electronic structures, phonon dynamics, and electrical conductivities of $Pb_{10}(PO_4)_6O$ and $Pb_9Cu(PO_4)_6O$ compounds, Arxiv, (2023).

[10] L. Liu, Z. Meng, X. Wang, H. Chen, Z. Duan, X. Zhou, H. Yan, P. Qin, Z. Liu, Semiconducting transport in $Pb_{10-x}Cu_x(PO_4)_6O$ sintered from $Pb_2SO_5$ and $Cu_3P$, Arxiv, (2023).

[11] H. Wu, L. Yang, J. Yu, G. Zhang, B. Xiao, H. Chang, Observation of abnormal resistance-temperature behavior along with diamagnetic transition in $Pb_{10-x}Cu_x(PO_4)_6O$-based composite, Arxiv, (2023).

[12] K. Kumar, N.K. Karn, V.P.S. Awana, Synthesis of possible room temperature superconductor LK-99: $Pb_9Cu(PO_4)_6O$, Superconductor Science and Technology, 36 (2023) 10LT02.

[13] H. Wu, L. Yang, B. Xiao, H. Chang, Successful growth and room temperature ambient-pressure magnetic levitation of LK-99, in, 2023.

[14] D. Garisto, Claimed superconductor LK-99 is an online sensation - but replication efforts fall short, Nature, 620 (2023) 253-253.

[15] D. Garisto, LK-99 isn't a superconductor - how science sleuths solved the mystery, Nature, 620 (2023) 705-706.